\input{aipcheck}

\documentclass[
    ,final            
  ]
  {aipproc}
\usepackage{graphicx}
\usepackage{amsfonts}
\usepackage{amssymb}
\usepackage{amsmath}
\usepackage{mathptmx}
\usepackage{rsfs}
\usepackage{bm}
\usepackage{ulem}
\usepackage{color}
\usepackage{shortvrb}

\layoutstyle{8x11single}

\begin{document}

\title{
Discriminating between two reformulations of SU(3) Yang-Mills theory on a lattice
}

\author{Akihiro Shibata}{
  address={Computing Research Center, High Energy Accelerator Research Organization (KEK), Tsukuba  305-0801, Japan}
}

\author{Kei-Ichi Kondo}{
  address={Department of Physics, Graduate School of Science, Chiba University, Chiba 263-8522, Japan}
}

\author{Seikou Kato}{
  address={Fukui National College of Technology, Sabae 916-8507, Japan}
}

\author{Toru Shinohara}{
  address={Department of Physics, Graduate School of Science, Chiba University, Chiba 263-8522, Japan}
}

\begin{abstract}
In oder to investigate quark confinement, we give a  new reformulation of the $SU(N)$ Yang-Mills theory on a lattice and present the results of the numerical simulations of the $SU(3)$ Yang-Mills theory on a lattice. 
The numerical simulations include the derivation of the linear potential for static interquark potential, i.e., non-vanishing string tension, in which the ``Abelian'' dominance and magnetic monopole dominance are established, confirmation of the dual Meissner effect by measuring the chromoelectric flux tube between quark-antiquark pair, the induced magnetic-monopole current, and the type of dual superconductivity, etc. 
\end{abstract}

\maketitle

\def\slash#1{\not\!#1}
\def\slashb#1{\not\!\!#1}
\def\slashbb#1{\not\!\!\!#1}


\section{Reformulation of lattice $SU(3)$ Yang-Mills theory} 

%
In the path-integral or functional-integral formulation, the basic ingredients are the action and the integration measure, by which the vacuum expectation value, say average) of an operator, is to be calculated. We can rewrite the original $SU(3)$ Yang-Mills action and the integration measure using either the \textbf{maximal option} or the \textbf{minimal option} \cite{KSM08} which includes the preceding works \cite{Cho80c,FN99a} as a special case. 
The resulting two reformulations written in terms of different variables are equivalent to each other, since each formulation corresponds to one of  the choices of the coordinates in the space of gauge field configurations. Therefore, we can use either reformulation (change of variables), instead of the original Yang-Mills theory.
The $SU(2)$ Yang-Mills theory was reformulated in \cite{KMS06,KMS05,Kondo06} using the field decomposition \cite{Cho80,DG79,FN99,Shabanov99} and the lattice version was also constructed \cite{KKMSSI06,IKKMSS07,SKKMSI07,KKS14}.
See \cite{KKSS14} for a review.

In what follows, we focus our studies on confinement of quarks in a defining representation, i.e., the \textbf{fundamental representation}. For this purpose, we use the \textbf{Wilson loop   average} for obtaining the \textbf{static quark potential}.
Remember that the Wilson loop operator is uniquely defined by specifying  a representation $R$, to which the source quark belongs. A remarkable fact is that the Wilson loop operator in the fundamental representation urges us to use the \textbf{minimal option} in the sense that it is exactly rewritten in terms of the field variables (i.e., the \textbf{color field} $n$ and the \textbf{restricted field} $V$) which identified with the field variables used to describe the minimal option.  This was shown in the process of deriving a \textbf{non-Abelian Stokes theorem for the Wilson loop operator} \cite{Kondo98b,KT99,Kondo99Lattice99,Kondo08} extending the original one \cite{DP89}. Therefore, the set of variables in the minimal option is a  natural and the best choice of coordinate in the space of gauge field configurations to describe the Wilson loop operator in the fundamental representation. At the same time, this fact tells us what is the dominant variable for the Wilson loop average. 

In view of this, we use the \textbf{reformulation of the $SU(3)$ Yang-Mills theory in the minimal option}  for  discussing confinement of quarks in the fundamental representation. 
Thus, the minimal option is superior to the maximal option for discussing confinement of quarks in the fundamental representation of the $SU(3)$ gauge group.
The reformulation of the lattice $SU(3)$ Yang-Mills theory in the minimal option is quickly reviewed as follows \cite{KSSMKI08,SKS10}.  
For the original $SU(3)$  gauge link variable $U_{x,\mu} \in SU(3)$,
we decompose it into the new variables $V_{x,\mu}$ and $X_{x,\mu}$ which have values in the $SU(3)$ group:
\begin{equation}
  SU(3) \ni U_{x,\mu}=X_{x,\mu}V_{x,\mu} , \quad X_{x,\mu}, V_{x,\mu} \in SU(3) .
\end{equation} 
Note that $V_{x.\mu}$ could be regarded as the dominant mode for quark confinement, while $X_{x,\mu}$ is the remainder. 
In this decomposition, we require that the restricted field  $V_{x,\mu}$ is transformed in the same way as the original gauge link variable $U_{x,\mu}$ and the remaining field $X_{x,\mu}$ as a site variable under the full $SU(3)$ gauge transformation $\Omega_{x}$:
\begin{subequations}
\begin{align}
 V_{x,\mu} & \longrightarrow V_{x,\mu}^{\prime} = \Omega_{x}V_{x,\mu} \Omega_{x+\mu}^{\dagger} , \quad \Omega_{x} \in G=SU(3)
\label{C35-V-transf}
\\
X_{x,\mu} & \longrightarrow X_{x,\mu}^{\prime} = \Omega_{x}X_{x,\mu} \Omega_{x}^{\dagger}  , \quad \Omega_{x} \in G=SU(3)
\label{C35-X-transf}
\end{align}
\label{C35-eq:gaugeTransf} 
\end{subequations}
for
\begin{align}
 U_{x,\mu} & \longrightarrow U_{x,\mu}^{\prime} = \Omega_{x}U_{x,\mu} \Omega_{x+\mu}^{\dagger}  , \quad \Omega_{x} \in G=SU(3) .
\end{align}

First, we introduce the key variable $\bm{h}_{x}$ called the \textbf{color field}. 
In the minimal option of $SU(3)$, a representation of the color field $\bm{h}_{x}$ is given by
\begin{equation}
 \bm{h}_{x}= \Theta_{x} \frac{\lambda^{8}}{2} \Theta_{x}^{\dagger}   \in Lie[SU(3)/U(2)] , \quad \Theta_{x}  \in SU(3), 
\end{equation}
with $\lambda^{8}$ being the Gell-Mann matrix for $SU(3)$ and $g_{x}$ the $SU(3)$ group element.
Once the color field $\bm{h}_{x}$ is introduced, the above decomposition is obtained by solving the (first) \textbf{defining equation}:
\begin{align}
   D_{\mu}^{(\epsilon)}[V]\bm{h}_{x}:= \epsilon^{-1} \left[  V_{x,\mu}\bm{h}_{x+\mu}-\bm{h}_{x}V_{x,\mu}\right]  =0 .
\label{C35-eq:def1}
\end{align}
In fact, this defining equation  can be solved exactly, and the solution is given by
\begin{subequations}
\label{C35-eq:decomp}%
\begin{align}
 X_{x,\mu} &= \widehat{L}_{x,\mu}^{\dag}\det(\widehat{L}_{x,\mu})^{1/3} \hat{g}_{x}^{-1},
\quad
 V_{x,\mu} = X_{x,\mu}^{\dag}U_{x,\mu} = \hat{g}_{x}\widehat{L}_{x,\mu}U_{x,\mu},
\\
 \widehat{L}_{x,\mu} &:= \left(  L_{x,\mu}L_{x,\mu}^{\dag}\right)^{-1/2}L_{x,\mu},
\\
 L_{x,\mu} &:= \frac{5}{3}\mathbf{1} + \sqrt{\frac{4}{3}}(\bm{h}_{x} + U_{x,\mu}\bm{h}_{x+\mu}U_{x,\mu}^{\dag}) + 8\bm{h}_{x}U_{x,\mu}\bm{h}_{x+\mu}U_{x,\mu}^{\dag} .
\end{align}
\label{C35-defeq}
\end{subequations}
Here the variable $\hat{g}_{x}$ is the $U(2)$ part which is undetermined from Eq.(\ref{C35-eq:def1}) alone.
In what follows, therefore, we put the second condition:
\begin{align}
 \hat{g}_{x} =1 ,
\label{C35-eq:def2}%
\end{align}
so that the above defining equations (\ref{C35-eq:def1}) and (\ref{C35-eq:def2}) correspond  respectively to the continuum version:
\begin{subequations}
\begin{align}
& \mathscr{D}_{\mu}[\mathscr{V}] \bm{h}(x) := \partial_\mu \bm{h}(x) -ig [\mathscr{V}_\mu(x), \bm{h}(x) ] =0, 
\\
& \mathscr{X}_{\mu}(x)-\frac43 [\bm{h}(x),[\bm{h}(x), \mathscr{X}_{\mu}(x)] = 0
 .
\end{align}
\label{C35-def-eq}
\end{subequations}
In the naive continuum limit, indeed, it is shown directly that (\ref{C35-defeq}) reproduces  the  decomposition in the continuum theory, which is obtained by solving (\ref{C35-def-eq}):
\begin{subequations}
\begin{align}
 \mathscr{A}_{\mu}(x) &= \mathscr{V}_{\mu}(x) + \mathscr{X}_{\mu}(x) ,
\nonumber\\
 \mathscr{V}_{\mu}(x)  &= \mathscr{A}_{\mu}(x)-\frac{4}{3}\left[ \bm{h}(x),\left[  \bm{h}(x),\mathscr{A}_{\mu}(x)\right]
\right]  -ig^{-1}\frac{4}{3}\left[  \partial_{\mu}\bm{h}%
(x), \bm{h}(x)\right]  ,
\\
 \mathscr{X}_{\mu}(x)  &  =\frac{4}{3}\left[ \bm{h}(x),\left[ \bm{h}(x),\mathscr{A}_{\mu}(x)\right]  \right]  +ig^{-1} \frac{4}{3}\left[  \partial_{\mu}\bm{h}(x),\bm{h}(x) \right]  .
\end{align}
\end{subequations}
Thus the decomposition is uniquely determined as   Eqs.(\ref{C35-eq:decomp}) up to the choice of $\hat{g}_{x}$ ($\ref{C35-eq:def2}$), once the color field $\bm{h}_{x}$ is specified.

In order to determine the configuration $\{ \bm{h}_{x}\}$ of color fields, we use the \textbf{reduction condition} which guarantees that  the new theory written in terms of  new variables  is equipollent to the original Yang-Mills  theory. 
Here, we use the reduction condition:
for a given configuration of the original link variables $\{ U_{x,\mu} \}$, a set of color fields $\left\{  \bm{h}_{x}\right\}  $ are obtained by minimizing the functional:
\begin{equation}
F_{\text{red}}[\{ \bm{h}_{x} \}]
= \sum_{x,\mu}\mathrm{tr}\left\{  (D_{\mu}^{(\epsilon)}[U]\bm{h}_{x})^{\dag}(D_{\mu}^{(\epsilon)}[U]\bm{h}_{x})\right\} .
 \label{C35-eq:reduction} 
\end{equation}
Consequently, the color field transforms under the gauge transformation as 
\begin{equation}
  \bm{n}_{x}  \rightarrow  \bm{n}_{x}^\prime = \Omega_{x} \bm{n}_{x} \Omega_{x}^{-1}  , \quad \Omega_{x} \in G=SU(3)
  .
  \label{C35-n-transf-2}
\end{equation}

\section{Restricted field dominance and magnetic monopole dominance} 

The lattice version of the Wilson loop operator $W_C[\mathscr{A}]$ is given by
\begin{align}
  W_C[U]  
:=  {\rm tr} \left[ \mathcal{P}  \prod_{<x,x+\mu> \in C}  U_{<x,x+\mu>}  \right]/{\rm tr}({\bf 1})  
 ,
\end{align}
where $\mathcal{P}$ is the path-ordered product. 
In the new formulation, we can define another non-Abelian Wilson loop operator $W_C[\mathscr{V}]$ by replacing the original Yang-Mills field $\mathscr{A}$ by the restricted field $\mathscr{V}$ in the original definition of the Wilson loop operator $W_C[\mathscr{A}]$. 
Similarly, the lattice version of the \textbf{restricted Wilson loop operator} $W_C[\mathscr{V}]$ is easily constructed as
\begin{align}
  W_C[V]  
:=  {\rm tr} \left[ \mathcal{P}  \prod_{<x,x+\mu> \in C}  V_{<x,x+\mu>}  \right]/{\rm tr}({\bf 1})  
 .
\end{align}
This is invariant under the gauge transformation (\ref{C35-V-transf}).

For $G=SU(3)$, the lattice version of the   \textbf{magnetic-monopole current} $K$  is given by using the restricted field $V$  as
\begin{subequations}
\begin{align}
   K_{x,\mu} 
=& \partial_{\nu}  {}^{\displaystyle *}\Theta_{x,\mu\nu}  
=\frac{1}{2}\epsilon_{\mu\nu\alpha\beta}\partial_{\nu}\Theta_{x,\alpha\beta}  ,
\\
 \epsilon^2 \Theta_{x,\alpha\beta} 
=&   {\rm arg} \Big[ {\rm tr} \Big\{ \left( \frac13 \bm{1} - \frac{2}{\sqrt{3}} \bm{n}_{x} \right) 
  V_{x,\alpha}V_{x+\alpha,\beta}V_{x+\beta,\alpha}^\dagger V_{x,\beta}^\dagger \Big\} \Big] ,
\\
&  V_{x,\alpha}V_{x+\alpha,\beta}V_{x+\beta,\alpha}^\dagger V_{x,\beta}^\dagger
= \exp\left(-ig \epsilon^2 \mathscr{F}_{\alpha\beta}[\mathscr{V}](x) \right)  .
\end{align}
\end{subequations}
The magnetic monopole current $K$ just defined in this way is gauge invariant. Indeed, it is easy to observe that $\Theta_{x,\mu\nu}$ is invariant under the gauge transformation (\ref{C35-n-transf-2}) and (\ref{C35-V-transf}),  and hence $K_{x,\mu}$ is also gauge-invariant. 
Then we can define the magnetic-monopole part of the Wilson loop operator by
\begin{align}
W_C[K] :=&   \exp \left( i  \sum_{x,\mu} K_{x,\mu} \Xi^{\Sigma}_{x,\mu} \right)
    ,
\nonumber\\
 \Xi^{\Sigma}_{x,\mu} :=& {\displaystyle\sum_{s^{\prime}}}
 \Delta_L^{-1}(s-s^{\prime})\frac{1}{2}\epsilon_{\mu\alpha\beta\gamma} 
\partial_{\alpha}S_{\beta\gamma}^{J}(s^{\prime}+\mu) , \quad
 \partial^\prime_{\alpha} S_{\alpha\beta}^{J}(x)=J_{\beta}(x) ,  
 \label{C35-WK}
\end{align}
where $\Xi_{x,\mu}$  is defined through the external source $J_{x,\mu}$ which is used to calculate the static potential,  $S_{\beta\gamma}^{J}(s^{\prime}+\mu)$ is a plaquette variable satisfying $\partial^\prime_{\beta} S_{\beta\gamma}^{J}(x)=J_{\gamma}(x)$ with the external source $J_{x,\mu}$ introduced to calculate the static potential, 
$\partial'$ denotes the backward lattice derivative
$\partial_{\mu}^{'}f_x=f_x-f_{x-\mu}$,  
$S^J_{x,\beta\gamma}$ denotes a surface bounded by the closed loop $C$ on which the electric source $J_{x,\mu}$ has its support, 
and $\Delta_L^{-1}(x-x')$ is the inverse Lattice Laplacian.

The static \textbf{quark-antiquark potential} $V(R)$ is obtained by taking the limit $T \rightarrow \infty$ from the Wilson loop average $\langle W_C[U]  \rangle$ for a rectangular loop $C=R \times T$.
In order to see the mechanism of quark confinement, we calculate three potentials:
\begin{enumerate}
\item
[(i)] the \textbf{full potential} $V_{\rm full}(R)$  calculated from the standard  $SU(3)$ Wilson loop average 
$\langle W_C[U] \rangle$:
\begin{align}
  V_{\rm full}(R) 
=& - \lim_{T \rightarrow \infty} \frac{1}{T} \ln 
\langle W_C[U] \rangle
 , 
 \label{C35-VfR}
\end{align}

\item
[(ii)] the \textbf{restricted potential} $V_{\rm rest}(R)$ calculated from the decomposed variable $\mathscr{V}$ through the restricted Wilson loop average $\langle W_C[V] \rangle$:
\begin{align}
  V_{\rm rest}(R) 
=& - \lim_{T \rightarrow \infty} \frac{1}{T} \ln \langle W_C[V] \rangle
 , 
 \label{C35-Vr}
\end{align}

\item
[(iii)] the \textbf{magnetic-monopole potential}  $V_{\rm mono}(R)$ calculated from the lattice counterpart (\ref{C35-WK}) of the continuum quantity   $\langle W_C[K] \rangle=\langle e^{  i (k, \Xi_{\Sigma})  }  \rangle$:
\begin{align}
V_{\rm mono}(R) 
=& - \lim_{T \rightarrow \infty} \frac{1}{T} \ln  \langle W_C[K] \rangle 
    ,
\end{align}

\end{enumerate}
Three potentials are gauge invariant quantities by construction.

\begin{figure}[t]
\includegraphics[scale=0.6]{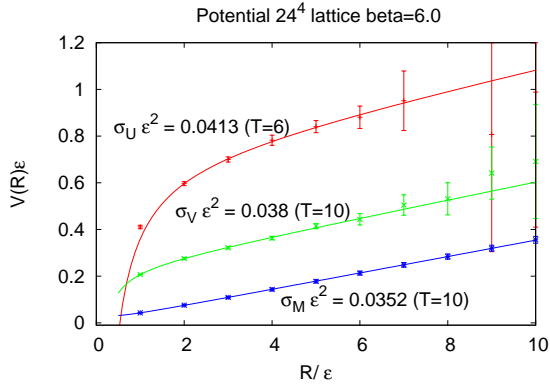}
\vspace{-0.5cm}
\caption{\cite{KSSK11} 
$SU(3)$ quark-antiquark potentials as functions of the quark-antiquark distance $R$: (from above to below) 
 (i) full potential $V_{\rm full}(R)$ (red curve),  (ii) restricted part $V_{\rm rest}(R)$ (green curve) and (iii) magnetic--monopole part  $V_{\rm mono}(R)$ (blue curve), measured at $\beta=6.0$ on $24^4$ using 500 configurations where $\epsilon$ is the lattice spacing.
}
\label{C35-fig:quark-potential}
\end{figure}

Numerical simulations are performed for $SU(3)$ Yang-Mills theory on the $24^4$ lattice according to the lattice reformulation explained above.

In Fig. \ref{C35-fig:quark-potential}, we compare the three quark-antiquark potentials (i), (ii) and (iii).  For each potential, we plot a set of point data for a specified value of $T$ (e.g., $T=6, 10$):
\begin{align}
    - \frac{1}{T} \ln \langle W_{C}[\cdot] \rangle \quad \text{versus} \quad R 
 , 
 \label{C35-VT}
\end{align}
and the curve represented by the function extrapolated to $T \rightarrow \infty$: 
\begin{align}
    V(R) =  \sigma R + b + c/R .
\end{align}

The results of our numerical simulations exhibit the \textbf{infrared restricted variable $\mathscr{V}$ dominance} in the string tension, e.g.,
\begin{equation}
\frac{\sigma_{\rm rest}}{\sigma_{\rm full}} = \frac{0.0380}{0.0413} \simeq 0.92,
\end{equation}
and the  \textbf{non-Abelian  magnetic monopole dominance} in the string tension, e.g.,
\begin{equation}
\frac{\sigma_{\rm mono}}{\sigma_{\rm full}} = \frac{0.0352}{0.0413} \simeq 0.85 .
\end{equation}
Thus, we have obtained the \textbf{infrared restricted variable $\mathscr{V}$ dominance} in the string tension  and the \textbf{non-Abelian  magnetic monopole dominance} in the string tension.
Both dominance are obtained in the gauge independent way.
See \cite{KSSK11} for more details.  
%


\section{Gauge-invariant chromoelectric field and flux tube formation}

\begin{figure}[ptb]
\includegraphics[
height=4cm,
]
{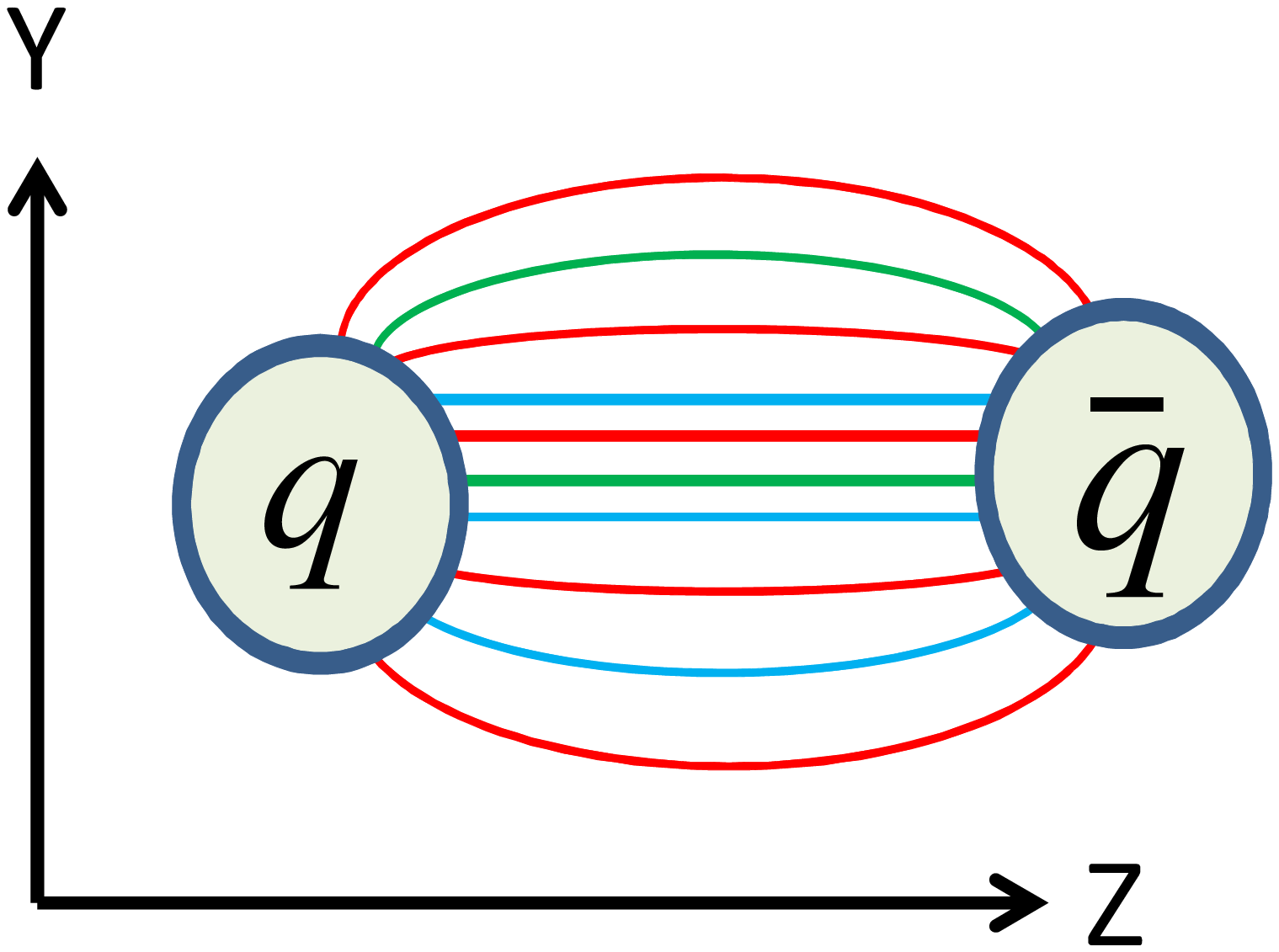} 
\quad\quad
\includegraphics[
origin=c,
height=4cm,
]
{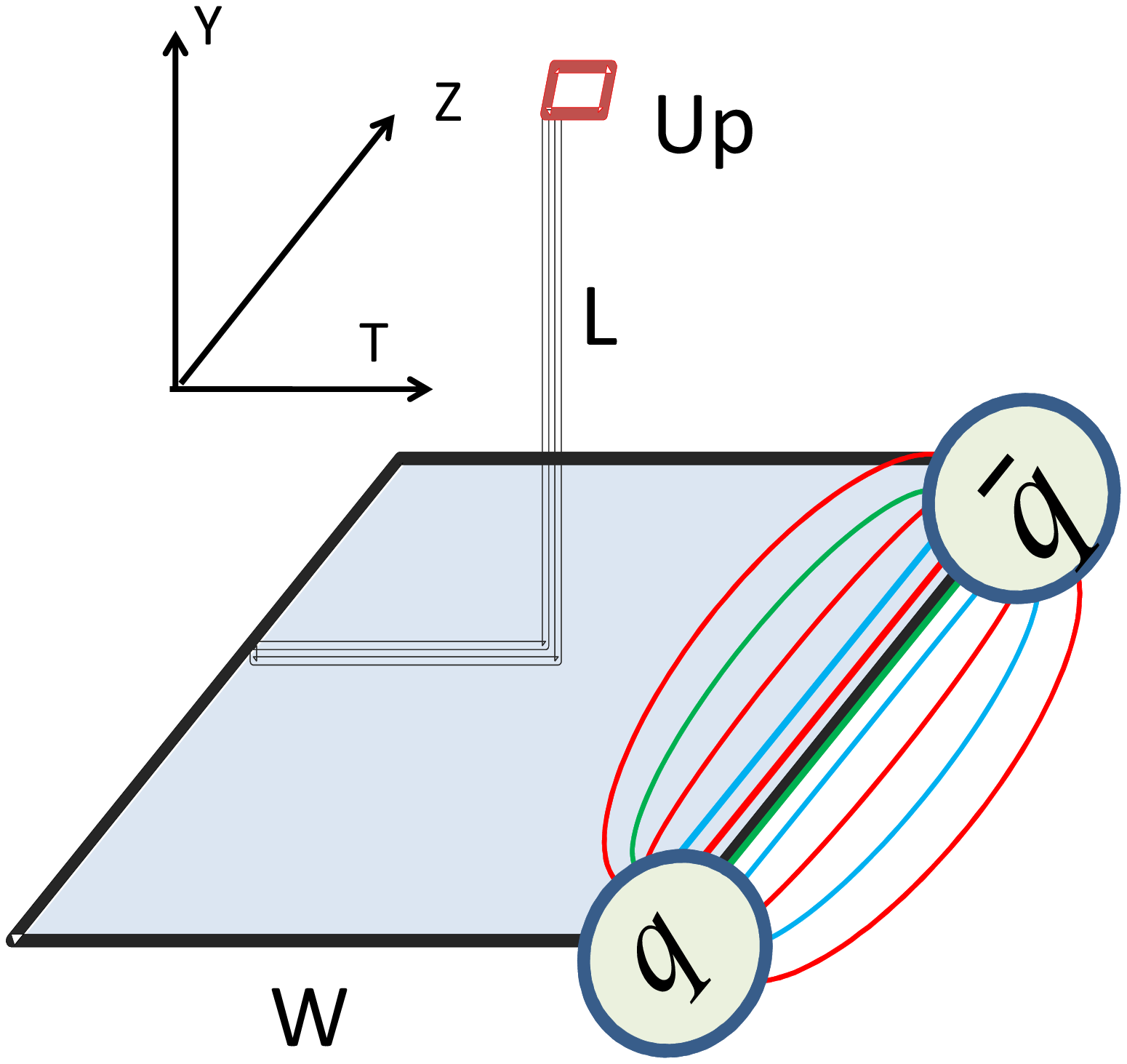}
\caption{(Left) 
The setup of measuring the chromo-flux produced by a quark--antiquark pair. 
(Right) 
The gauge-invariant connected correlator ($U_{p}LWL^{\dag})$ between a plaquette $U$ and the Wilson loop $W$.
}%
\label{C35-fig:Operator}%
\end{figure}

In order to extract the chromo-field, we use a gauge-invariant correlation function proposed by Di Giacomo, Maggiore and Olejnik  \cite{GMO90}.  
The chromo-field created by a quark-antiquark pair in $SU(N)$ Yang-Mills theory is measured by using a gauge-invariant connected correlator between a plaquette and the Wilson loop  
(see Fig.\ref{C35-fig:Operator}):%
\begin{equation}
\rho_{_{U_P}}:=\frac{\left\langle \mathrm{tr}\left(  U_{P}L^{\dag}WL\right)
\right\rangle }{\left\langle \mathrm{tr}\left(  W\right)  \right\rangle
}-\frac{1}{N}\frac{\left\langle \mathrm{tr}\left(  U_{P}\right)
\mathrm{tr}\left(  W\right)  \right\rangle }{\left\langle \mathrm{tr}\left(
W\right)  \right\rangle }, 
\label{C35-eq:Op}%
\end{equation}
where $W$ is the Wilson loop in
$Z$-$T$ plane representing a pair of quark and antiquark, $U_{P}$ a plaquette variable as the probe operator to measure the chromo-field strength at the point $P$, and $L$ the Wilson line connecting the source $W$ and the probe $U_{P}$.
Here $L$ is necessary to guarantee the gauge invariance of the correlator $\rho_{_{U_P}}$ and hence the probe is identified with $LU_PL^\dagger$. 
The symbol $\left\langle \mathcal{O}\right\rangle $ denotes the average of the operator $\mathcal{O}$ in the space and the ensemble of the configurations. 
In the naive continuum limit $\epsilon \to 0$, indeed,  $\rho_{_{U_P}}$ reduces to the field strength in the presence of the $q\bar q$ source:
\begin{equation}
\ \rho_{_{U_P}}%
\overset{\varepsilon\rightarrow0}{\simeq}g\epsilon^{2}\left\langle
\mathscr{F}_{\mu\nu}\right\rangle _{q\bar{q}}:=\frac{\left\langle
\mathrm{tr}\left(  ig\epsilon^{2}  \mathscr{F}_{\mu\nu}L^{\dag}WL \right)
\right\rangle }{\left\langle \mathrm{tr}\left(  W\right)  \right\rangle
}+O(\epsilon^{4}) ,
\end{equation}
where we have used $U_{x,\mu}=\exp (-ig\epsilon \mathscr{A}_\mu(x))$ and hence $U_{P}= \exp(-ig\epsilon^2 \mathscr{F}_{\mu\nu} )$.
Thus, the \textbf{gauge-invariant chromo-field strength} $F_{\mu\nu}[U]$ produced by a $q\bar q$ pair is given by 
\begin{equation}
F_{\mu\nu}[U] := \epsilon^{-2} \sqrt{\frac{\beta}{2N}}\rho_{_{U_P}} ,
\end{equation}
where $\beta:=2N/g^2$ is the lattice gauge coupling constant. 
Note that the connected correlator $\rho_{_{U_P}}$ is sensitive to the field strength, while  the disconnected one probes the squared field strength:
\begin{equation}
\rho_{_{U_P}}^{\prime}:=\frac{\left\langle \mathrm{tr}\left(  W\right)
\mathrm{tr}\left(  U_{P}\right)  \right\rangle }{\left\langle \mathrm{tr} 
\left(  W\right)  \right\rangle }-\left\langle \mathrm{tr}\left( U_{P}\right)  \right\rangle \overset{\varepsilon\rightarrow0}{\simeq} 
g\epsilon^{4}\left[  \left\langle \mathscr{F}_{\mu\nu}^{2}\right\rangle_{q\bar{q}} - \left\langle \mathscr{F}_{\mu\nu}^{2}\right\rangle _{0} \right]  .
\end{equation}

\begin{figure}[ptb]
\includegraphics[
scale=0.30,
angle=270,
]
{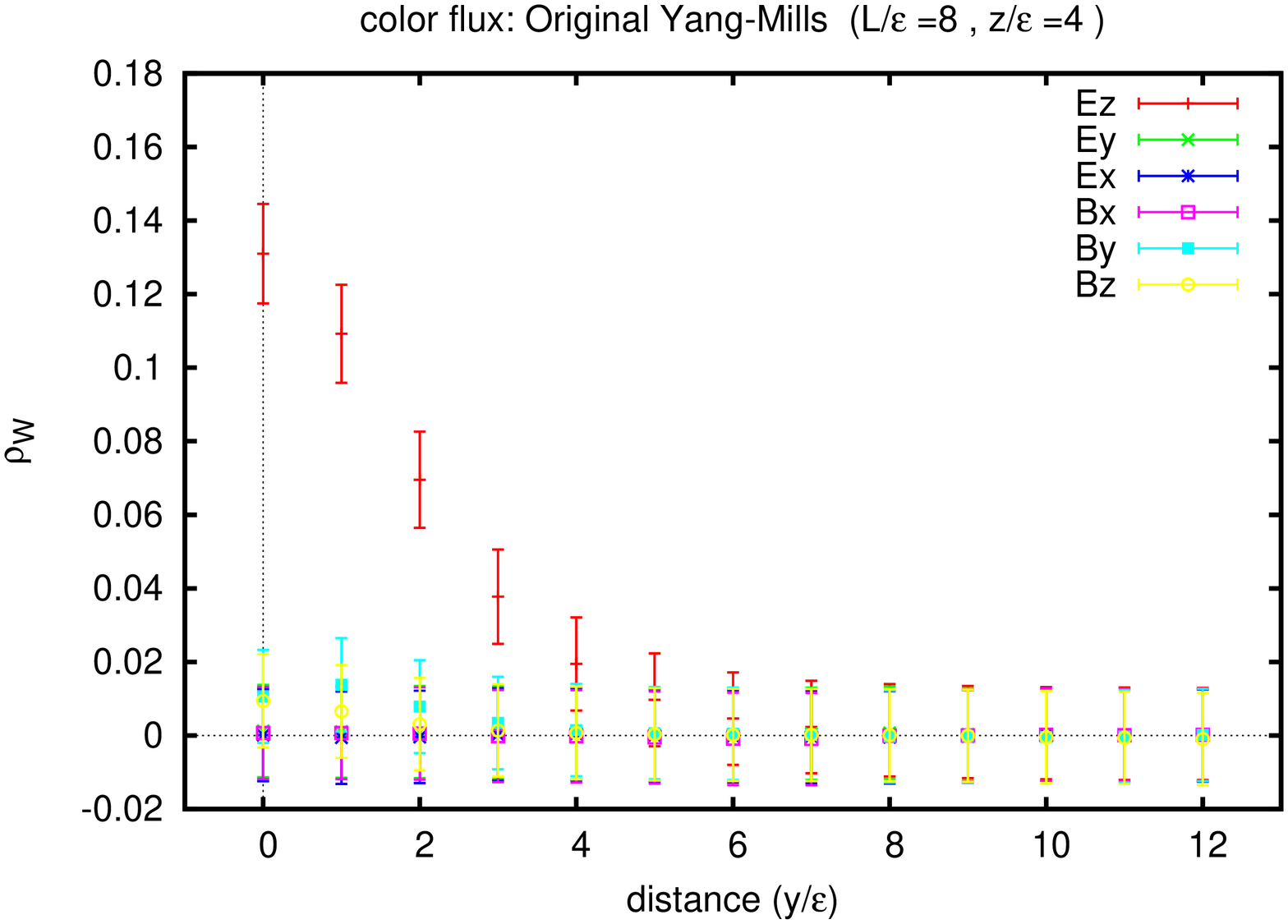}  
\ 
\includegraphics[
scale=0.30,
angle=270,
]
{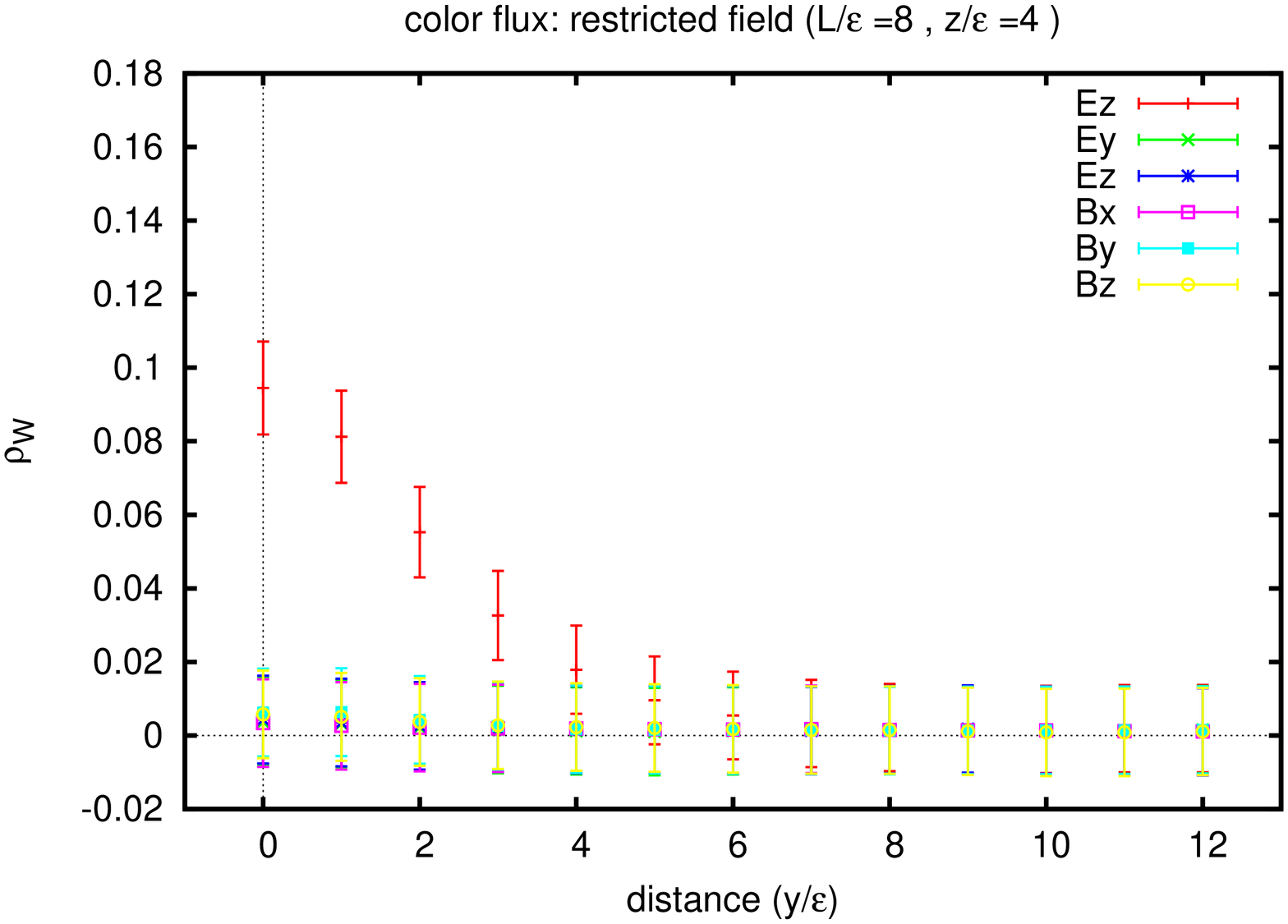}
\caption{
\cite{SKKS13} 
Measurement of components of the chromoelectric field $\bm{E}$ and chromomagnetic field  $\bm{B}$ as functions of the distance $y$ from the $z$ axis. 
(Left panel) the original $SU(3)$ Yang-Mills field, 
(Right panel) the restricted  field. 
}%
\label{C35-fig:measure}%
\end{figure}

We measure correlators between the plaquette $U_P$ and the chromo-field strength of the restricted field $V_{x,\mu}$ as well as the original Yang-Mills field $U_{x,\mu}$.
See the left panel of Fig.~\ref{C35-fig:Operator}.
Here the quark and antiquark source is introduced as $8\times8$ Wilson loop ($W$) in the $Z$-$T$ plane, and the probe $(U_{p})$ is set at the center of the Wilson loop and moved along the $Y$-direction. 
The left and right panel of Fig.~\ref{C35-fig:measure} show respectively the results of measurements for the chromoelectric and chromomagnetic fields $F_{\mu\nu}[U]$ for the original  $SU(3)$ field $U$ and $F_{\mu\nu}[V]$ for the restricted  field $V$, where the field strength $F_{\mu\nu}[V]$  is obtained by using $V_{\,x,\mu}$ in  (\ref{C35-eq:Op}) instead of $U_{x,\mu}$:
\begin{equation}
F_{\mu\nu}[V] := \sqrt{\frac{\beta}{2N}} \tilde\rho_{_{V_P}} , \quad 
\tilde\rho_{_{V_P}} := \frac{\left\langle \mathrm{tr}\left(   V_{P}L^{\dag}WL \right)
\right\rangle }{\left\langle \mathrm{tr}\left(  W\right)  \right\rangle
}-\frac{1}{N}\frac{\left\langle \mathrm{tr}\left(  V_{P}\right)
\mathrm{tr}\left(  W\right)  \right\rangle }{\left\langle \mathrm{tr}\left(
W\right)  \right\rangle } .  
\label{C35-cf1-5-SU3}
\end{equation}
We have checked that even if  $W[U]$ is replaced by $W[V]$, together with replacement of the probe $LU_{P}L^\dagger$ by the corresponding $V$ version, the  change in the magnitude of the field strength $F_{\mu\nu}$ remains within at most a few \%.

From Fig.\ref{C35-fig:measure} we find that  only the $E_{z}$ component of the \textbf{chromoelectric field} $(E_x,E_y,E_z)=(F_{10},F_{20},F_{30})$ connecting $q$ and $\bar q$ has non-zero value for both the restricted field $V$ and the original Yang-Mills field $U$.
The other components are zero consistently within the numerical errors. 
This means that the chromomagnetic field $(B_x,B_y,B_z)=(F_{23},F_{31},F_{12})$ connecting $q$ and $\bar q$ does not exist  and that the chromoelectric field is parallel to the $z$ axis on which  quark and antiquark are located.
The magnitude $E_{z}$ quickly decreases in the distance $y$ away from the Wilson loop.

\begin{figure}[ptb]
\includegraphics[
height=7.0cm,
angle=270
]
{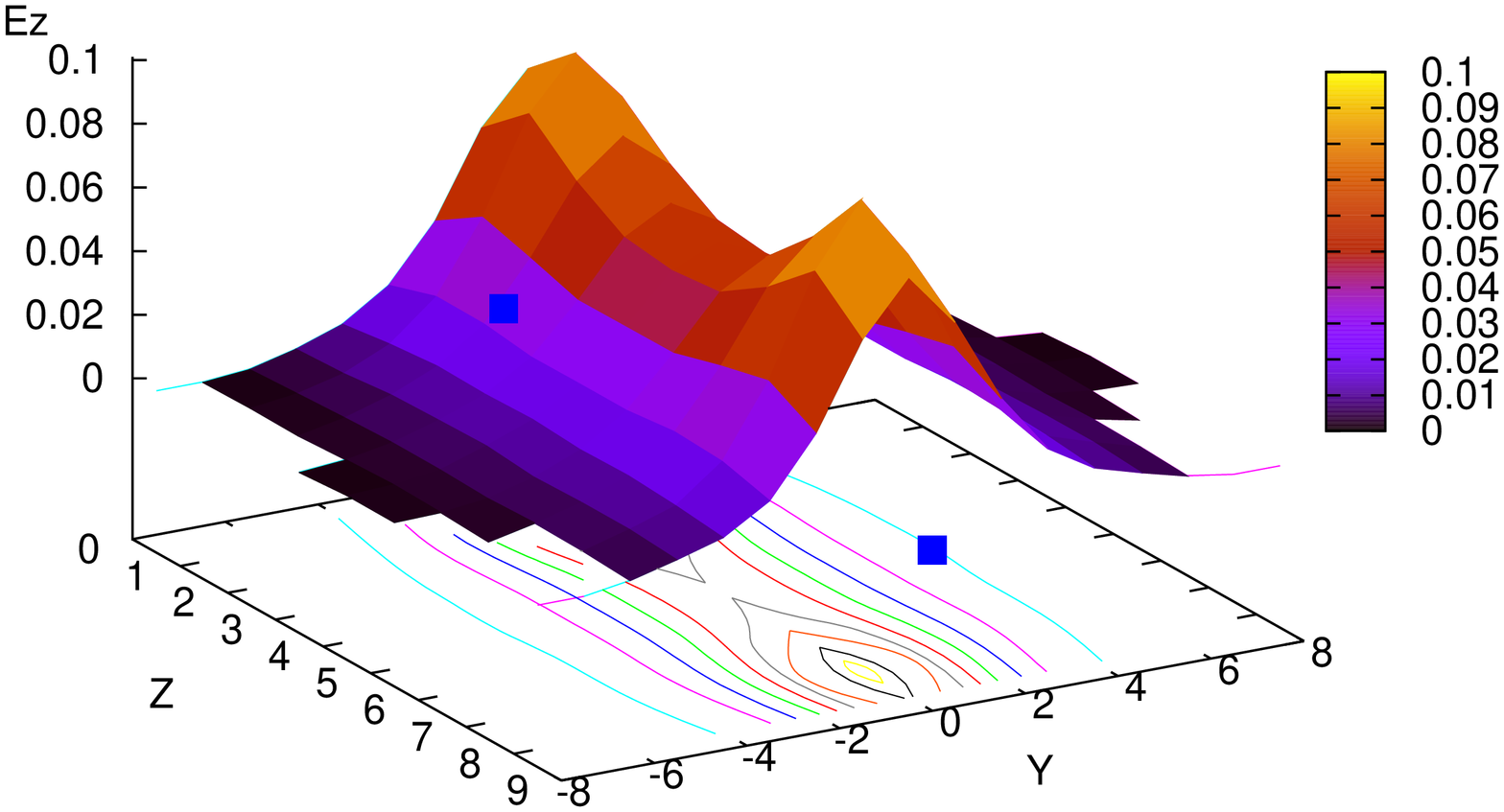} 
\quad
\includegraphics[
height=7.0cm,
angle=270
]
{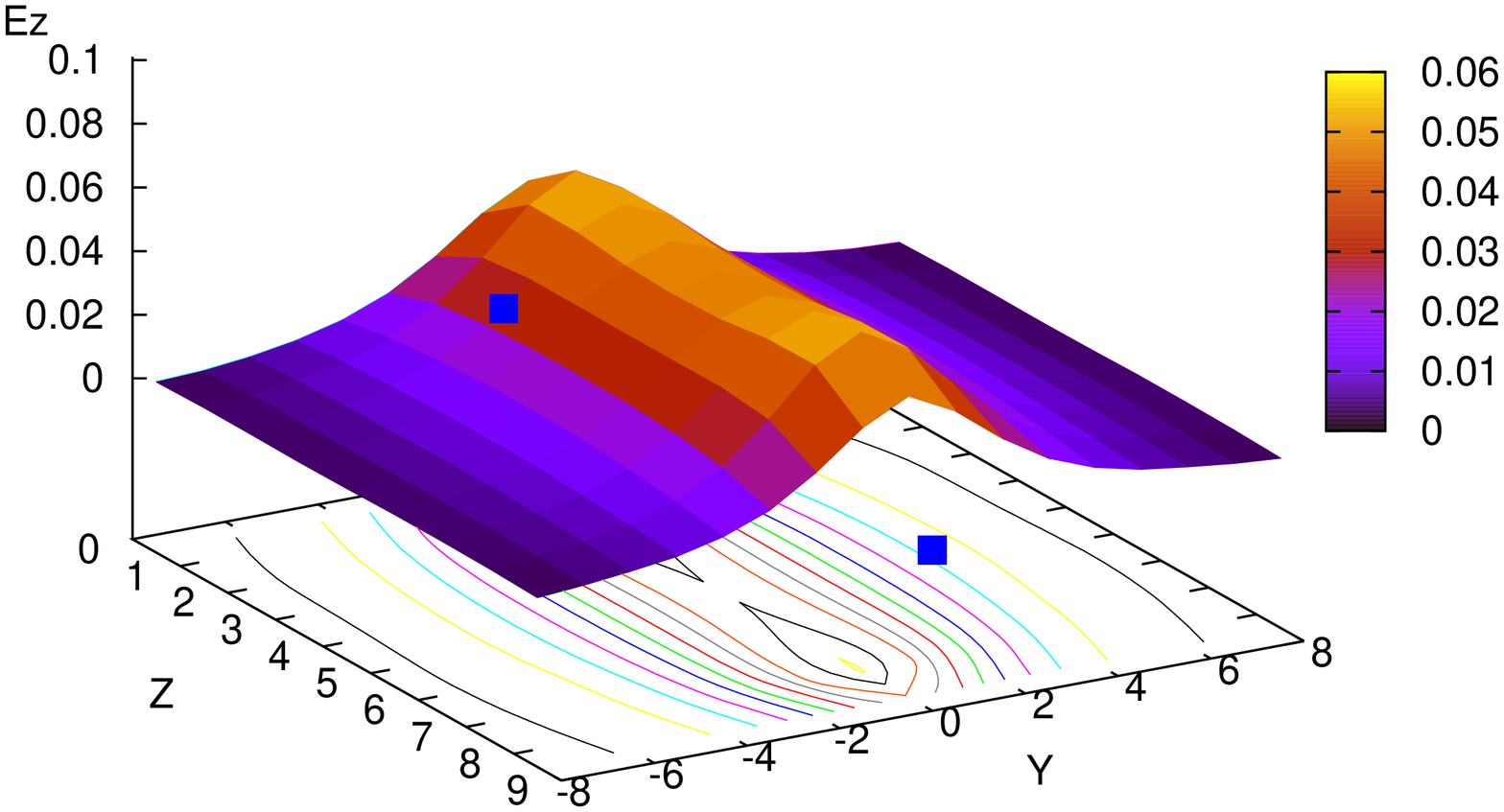} 
\caption{\cite{SKKS13}
The distribution in $Y$-$Z$ plane of the chromoelectric field $E_z$ connecting a pair of quark and antiquark: 
(Left panel) chromoelectric field produced from the original Yang-Mills field, 
(Right panel) chromoelectric field produced from the restricted  field. 
}%
\label{C35-fig:fluxtube}%
\end{figure}

To see the profile of the non-vanishing component $E_z$ of the chromoelectric field in detail, we explore the distribution of chromoelectric field on the 2-dimensional plane. 
 Fig.~\ref{C35-fig:fluxtube} shows the distribution of $E_{z}$ component of the chromoelectric field, where the quark-antiquark source represented as $9\times11$ Wilson loop $W$ is placed at $(Y,Z)=(0,0), (0,9)$, and the probe $U$ is displaced on the $Y$-$Z$ plane at the midpoint of the $T$-direction. 
 The position of a quark and an antiquark is marked by the solid (blue) box. The magnitude of $E_{z}$ is shown by the height of the 3D plot and also the contour plot in the bottom plane.
 The left panel of Fig.~\ref{C35-fig:fluxtube} shows the plot of $E_z$ for the $SU(3)$ Yang-Mills field $U$, and the right panel of Fig.~\ref{C35-fig:fluxtube} for the restricted field $V$. 
We find that the magnitude $E_{z}$ is quite uniform for the restricted part $V$, while it is almost uniform for the original part $U$ except for the neighborhoods of the locations of $q$, $\bar q$ source. 
This difference is due to  the contributions from the remaining part $X$ which affects only the  short distance, as will be discussed later.

\section{Magnetic current and dual Meissner effect for $SU(3)$ case}

Next, we investigate the relation between the chromoelectric flux and the  magnetic current. 
The magnetic(-monopole) current can be calculated as
\begin{equation}
 k= \delta{}^{\displaystyle *}F[V] ={}^{\displaystyle *}d F[V] ,
\label{C35-def-k2}
\end{equation}
where $F[V]$ is the field strength (\ref{C35-cf1-5-SU3}) defined from the the restricted field $V$  in the presence of the $q\bar q$ source,
$d$
the exterior derivative, $\delta$ codifferential, and $^{\ast}$ denotes the Hodge dual operation. 
Note that non-zero magnetic current follows from violation of the Bianchi identity  
(If the field strength was given by the exterior derivative of some  field $A$ (one-form), $ F=dA$, \ we would obtain $k=\delta{}^{\displaystyle *}F={}^{\displaystyle *}d^{2}A=0$).

\begin{figure}[ptb]
\includegraphics[
scale=1.0
]
{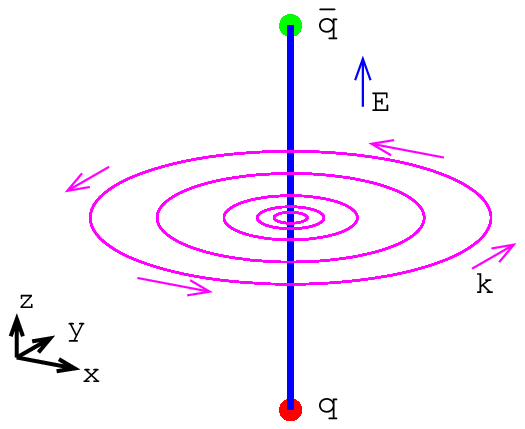} 
\quad
\includegraphics[
scale=0.6
]
{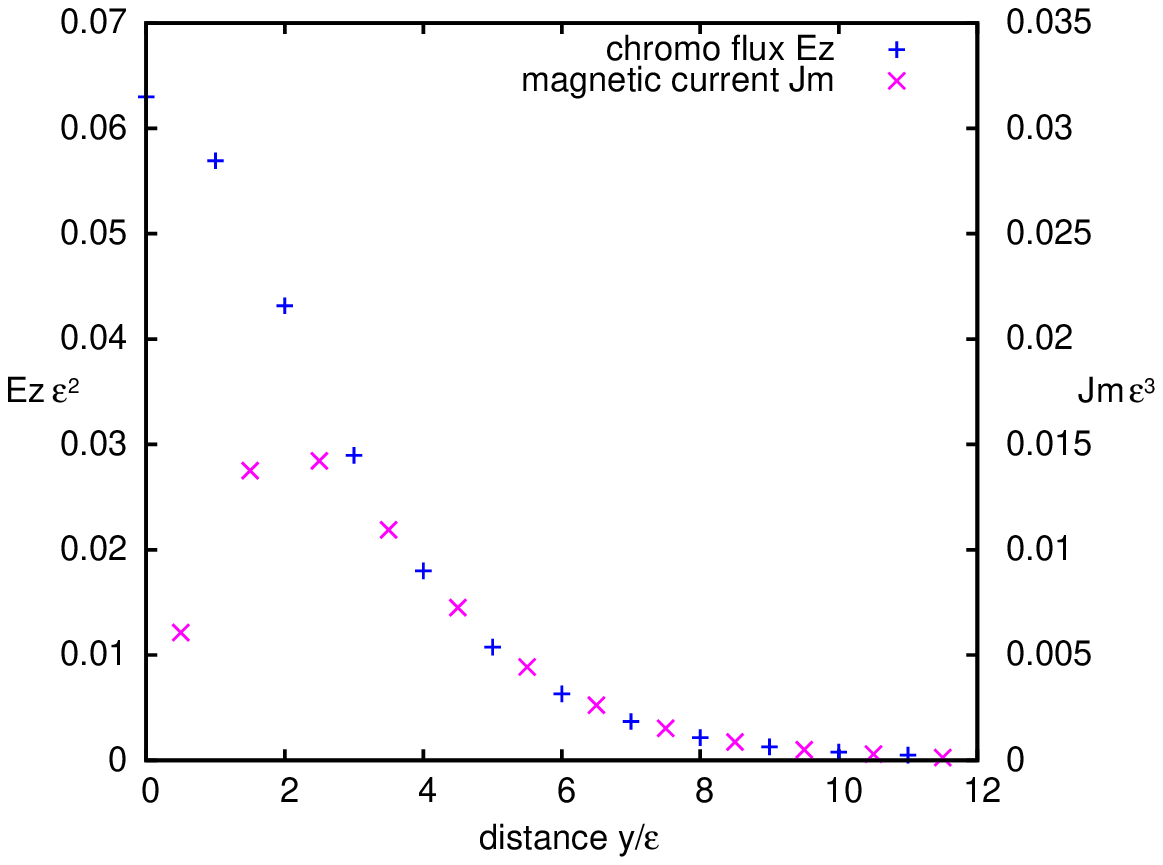} 
\caption{\cite{SKKS13}{}
The magnetic-monopole current $\mathbf{k}$ induced around the flux along the $z$ axis connecting a  quark-antiquark pair.
(Left panel) The positional relationship between the chromoelectric field $E_{z}$ and the magnetic current $\mathbf{k}$. 
(Right panel) The magnitude of the chromo-electronic current $E_{z}$ and the magnetic current  $J_{m}=|\mathbf{k}|$ as functions of the distance $y$ from the $z$ axis. 
}
\label{C35-fig:Mcurrent-SU3}%
\end{figure} 

Fig.~\ref{C35-fig:Mcurrent-SU3} shows the  magnetic current measured in $X$-$Y$ plane at the midpoint of quark and antiquark pair in the $Z$-direction. 
The left panel of Fig.~\ref{C35-fig:Mcurrent-SU3} shows the positional relationship between chromoelectric flux and  magnetic current.
The right panel of Fig.~\ref{C35-fig:Mcurrent-SU3} shows the magnitude of the  chromoelectric field $E_z$ (left
scale) and the magnetic current $k$ (right scale). 
The existence of non-vanishing  \textbf{magnetic current} $k$ around the  \textbf{chromoelectric field} $E_z$ supports the  \textbf{dual superconductivity} which is the dual picture of the ordinary superconductor exhibiting the electric current $J$ around the magnetic field $B$.

In our formulation, it is possible to define a gauge-invariant magnetic-monopole current  $k_{\mu}$  by using $V$-field,
which is obtained from the field strength $\mathscr{F}[\mathscr{V}]$ of the restricted field $\mathscr{V}$, as suggested from the non-Abelian Stokes theorem.
It should be also noticed that this
magnetic-monopole current  is a non-Abelian magnetic monopole extracted from the $V$ field, which corresponds to the maximal stability group $\tilde{H}=U(2)$.
The magnetic-monopole current  $k_{\mu}$ defined in this way can be used to study the magnetic current around the chromoelectric flux tube, instead of the above definition of $k$  (\ref{C35-def-k2}).
The comparison of two monopole currents $k$ is to be done in the future works. 

These are numerical evidences supporting  {``non-Abelian'' dual superconductivity due to non-Abelian  magnetic monopoles   as a mechanism for quark confinement in SU(3) Yang-Mills theory}.

\section{Type of dual superconductivity}

\begin{figure}[ptb]
\includegraphics[
width=5.0cm,
angle=270
]
{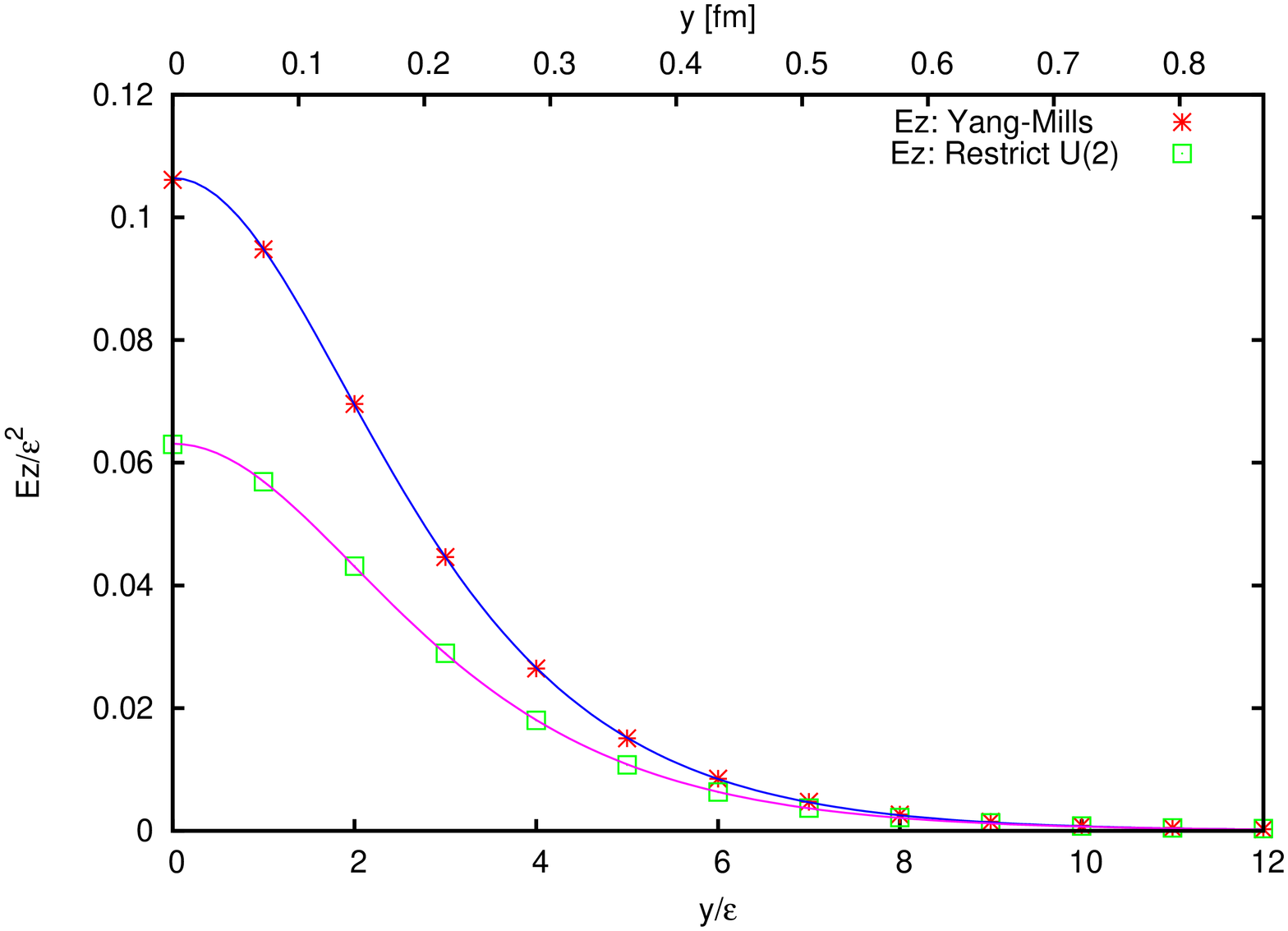} 
\includegraphics[,
width=5.0cm,
angle=270
]
{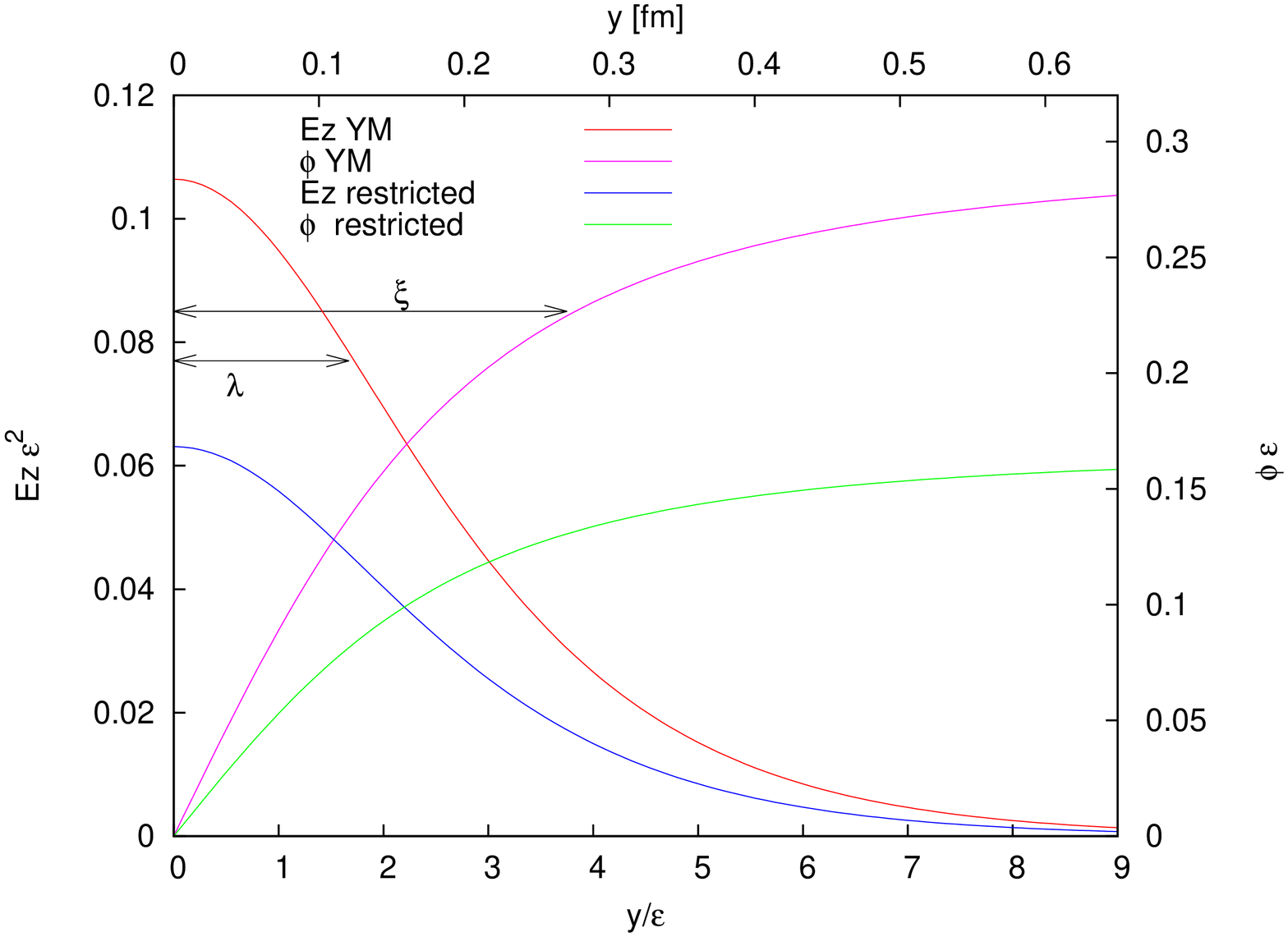}  
\caption{\cite{SKKS13} 
(Left panel)
The plot of the chromoelectric field $E_z$ versus the distance $y$ in units of the lattice spacing $\epsilon$ and the fitting   as a function $E_z(y)$ of  $y$ according to (\ref{C35-Clem-fit}). 
The red cross for the original $SU(3)$ field and the green square symbol for the restricted field.  
(Right panel) The order parameter $\phi$ reproduced as a function $\phi(y)$ of  $y$ according to (\ref{order-f}), together with the chromoelectric field $E_z(y)$.
}
\label{C35-fig:type}%
\end{figure}

Moreover, we investigate the QCD vacuum, i.e., type of the dual superconductor. 
The left panel of Fig.\ref{C35-fig:type} is the plot for the chromoelectric field $E_z$ as a function of the distance $y$ in units of the lattice spacing $\epsilon$ for the original $SU(3)$ field and for the restricted  field.

In order to examine the \textbf{type of the dual superconductivity}, we apply the formula for the magnetic field  derived by Clem \cite{Clem75} in the ordinary superconductor based on the \textbf{Ginzburg-Landau (GL) theory} to the chromoelectric field in the dual superconductor.
In the GL theory, the gauge field $A$ and the scalar field $\phi$ obey simultaneously  the GL equation:
\begin{equation}
 (\partial^\mu -iq A^\mu)(\partial_\mu -iq A_\mu) \phi + \lambda (\phi^* \phi - \eta^2) = 0 ,
\end{equation}
and the Ampere equation:
\begin{equation}
 \partial^\nu F_{\mu\nu} + iq [\phi^* (\partial_\mu \phi -iq A_\mu \phi)  - (\partial_\mu \phi -iq A_\mu \phi)^* \phi] = 0 .
\end{equation}

Usually, in the dual superconductor of the type II, it is justified to use the asymptotic form $K_0(y/\lambda)$ to fit the chromoelectric field in the large $y$ region (as the solution of the Ampere equation in the dual GL theory).  
However, it is clear that this solution cannot be applied to the small $y$ region, as is easily seen from the fact that $K_0(y/\lambda) \to \infty$ as $y \to 0$. 
In order to see the difference between type I and type II, it is crucial to see the relatively small $y$ region.
Therefore, such a simple form cannot be used to detect the type I dual  superconductor. 
However, this important aspect was ignored in the preceding studies except for a work \cite{Cea:2012qw}.

On the other hand, Clem \cite{Clem75} does not obtain the analytical solution of the GL equation explicitly and use an approximated form for the scalar field  $\phi$ (given below in (\ref{order-f})).
This form is used to solve the Ampere equation exactly to obtain the analytical form for the gauge field $A_\mu$ and the resulting magnetic field $B$.
This method does not change the behavior of the gauge field in the long distance, but it gives a finite value for the gauge field even at the origin. 
Therefore, we can obtain the formula which is valid for any distance (core radius) $y$ from the  axis connecting $q$ and $\bar{q}$: the profile of chromoelectric field in the dual superconductor is obtained:
\begin{equation}
E_{z}(y)=\frac{\Phi}{2\pi}\frac{1}{\zeta\lambda}\frac{K_{0}(R/\lambda)}%
{K_{1}(\zeta/\lambda)},\text{ }R=\sqrt{y^{2}+\zeta^{2}} ,
\label{C35-Clem-fit}%
\end{equation}
provided that the scalar field is given by (See the right panel of Fig.\ref{C35-fig:type})
\begin{equation}
 \phi(y) = \frac{\Phi}{2\pi} \frac{1}{\sqrt{2}\lambda} \frac{y}{\sqrt{y^2+\zeta^2}} ,
 \label{order-f}
\end{equation}
where $K_{\nu}$ is the modified Bessel function of the $\nu$-th order,
$\lambda$ the parameter corresponding to the London \textbf{penetration length}, $\zeta$
a variational parameter for the core radius, and $\Phi$ external electric flux. 
In the dual superconductor, we define the \textbf{GL parameter} $\kappa$ as the ratio of the London penetration length $\lambda$ and the \textbf{coherence length} $\xi$ which measures the coherence of the magnetic monopole condensate (the dual version of the Cooper pair condensate):
\begin{equation}
\kappa= \frac{\lambda}{\xi}  \ .
\end{equation}  
It is given by \cite{Clem75}
\begin{equation}
\kappa=\sqrt{2} \frac{\lambda}{\zeta} \sqrt{1-K_{0}^{2}(\zeta/\lambda)/K_{1}^{2}(\zeta/\lambda)} .
\end{equation}

\begin{table}[t] 
\fontsize{6.5pt}{0pt}\selectfont
\begin{tabular}{|l||c|c|c|| c|c|c|c|c|}\hline
& $a\epsilon^2$    & $b\epsilon$    & $c$       & $\lambda/\epsilon$ 
&$ \zeta/\epsilon$ & $\xi/\epsilon$ &  $\Phi$ & $\kappa$ \\
\hline
SU(3) Yang-Mills field 
& $0.804 \pm 0.04$ & $0.598\pm 0.005$ & $1.878\pm 0.04$ & $1.672 \pm 0.014$ 
& $3.14\pm 0.09$ &  $3.75 \pm 0.12$  & $4.36 \pm 0.3$& $0.45 \pm 0.01$ \\
\hline
restricted field 
& $0.435 \pm 0.03 $ & $0.547 \pm 0.007$ & $1.787 \pm 0.05 $ & $1.828 \pm 0.023$ 
& $3.26 \pm 0.13$    & $3.84 \pm 0.19$    &$2.96 \pm 0.3 $     & $0.48 \pm 0.02$ \\
\hline
\end{tabular}
\caption{
The properties of the Yang-Mills vacuum as the dual superconductor obtained by fitting the data of chromoelectric field with the prediction of the dual Ginzburg-Landau theory.  
}
\label{C35-Table:GL-fit}
\end{table}

See Fig.\ref{C35-fig:type}.
Our data clearly shows that the dual superconductor of $SU(3)$ Yang-Mills theory is \textbf{type I} with 
\begin{equation}
 \kappa=0.45 \pm 0.01 .
\end{equation} 
This result is consistent with a quite recent result obtained independently by  Cea, Cosmai and Papa \cite{Cea:2012qw}. The London penetration length $\lambda=0.1207(17)$fm and the coherence length $\xi=0.2707(86)$fm is obtained in units of the string tension $\sigma_{\text{phys}}=(440\text{MeV})^2$, and data of lattice spacing is taken from the Table I in Ref.\cite{Edward98}. 
Moreover, our result shows that the restricted  part plays the dominant role in determining the type of the non-Abelian dual superconductivity of the $SU(3)$ Yang-Mills theory, 
 i.e., type I with  
\begin{equation}
 \kappa=0.48 \pm 0.02 ,
\end{equation} 
$\lambda=0.132(3)$fm and $\xi=0.277(14)$fm.
This is a novel feature overlooked in the preceding studies. 
Thus the \textbf{restricted-field dominance} can be seen also in the determination of the type of dual superconductivity where the discrepancy is just the normalization of the chromoelectric field at the core $y=0$, coming from the difference of the total flux $\Phi$. 
These are  gauge-invariant results. 
Note again that this restricted-field and the non-Abelian magnetic monopole extracted from it reproduce the string tension in the static quark--antiquark potential.

Our result should be compared with the result obtained by using the Abelian projection:  Matsubara et. al \cite{Matsubara:1993nq} suggests $\kappa=0.5 \sim 1$(which is $\beta$ dependent), border of type I and type II for both $SU(2)$ and $SU(3)$. 
In $SU(2)$ case, on the other hand, there are other works \cite{Suzuki:2009xy,Chernodub:2005gz} which conclude that the type of vacuum is at the border of type I and type II.
Our results \cite{KKS14} are consistent with the border of type I and type II for the  $SU(2)$ Yang-Mills theory on the lattice, as already shown in the above.


We should mention the work \cite{Cardoso:2010kw} which concludes that  the dual superconductivity of $SU(3)$ Yang-Mills theory is type II with $\kappa=1.2 \sim 1.3$.
This conclusion seems to contradict our result for $SU(3)$. 
If the above formula (\ref{C35-Clem-fit}) is applied to the data of \cite{Cardoso:2010kw}, we have the same conclusion, namely, the type I with $\kappa=0.47 \sim 0.50$. 
Therefore, the data obtained in \cite{Cardoso:2010kw} are consistent with ours. 
The difference between type I and type II is attributed to the way of fitting the data with the formula for the chromo-field.

\section{Color direction field and color symmetry}

\begin{figure}[t]
\includegraphics[scale=0.40]{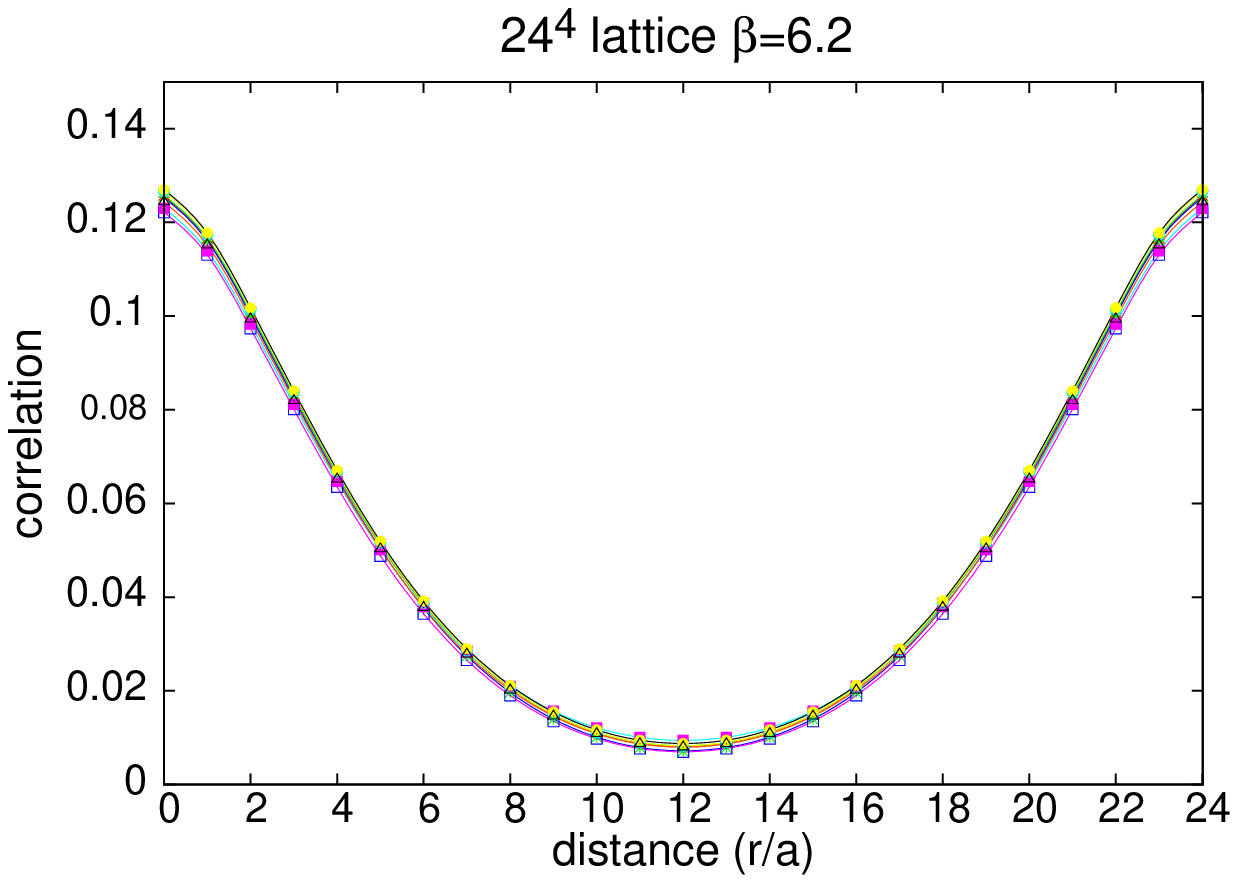}
\quad\quad
\includegraphics[scale=0.40]{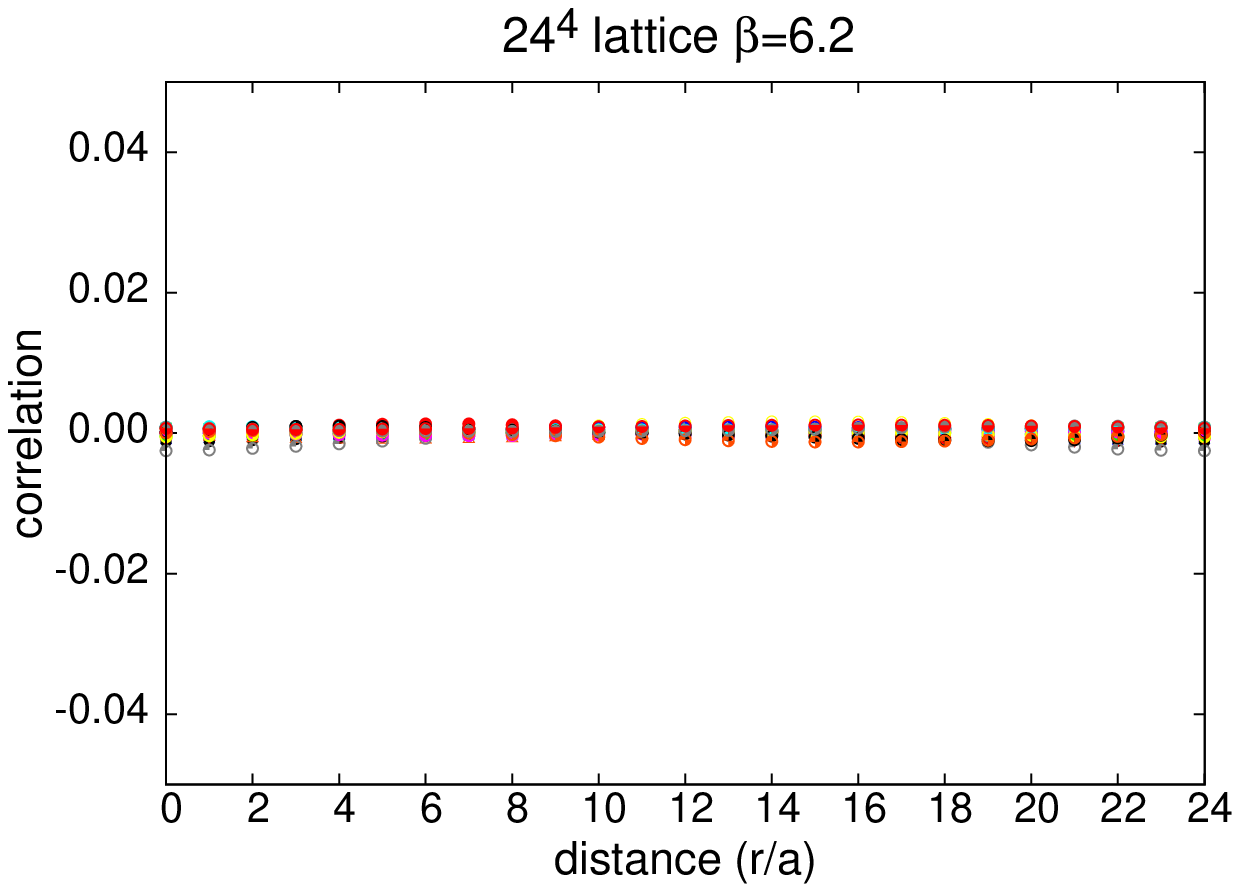}
\vspace{-0.5cm}
\caption{\cite{KSSK11} 
Color field correlators $\langle  n^A(x) n^B(0) \rangle$ ($A,B=1, \cdots, 8$) as functions of the distance $r:=|x|$ measured at  $\beta=6.2$ on   24$^4$ lattice,  using 500 configurations under the Landau gauge. 
(Left panel) $A=B$, 
(Right panel) $A \not= B$.
}
\label{C35-fig:color-field-corr}
\end{figure}

Fig.\ref{C35-fig:color-field-corr} shows  two-point correlation functions of color field $\langle  n^A(x) n^B(0) \rangle$ versus the distance $r:=|x|$. 
All plots of correlators for $A=B=1,2, \cdots, 8$ overlap on top of each other, and hence they can be fitted by a common non-vanishing  function $D(r)$ (left panel), while all correlators for $A \not= B$ are nearly equal to zero (right panel). 
Therefore,  the correlators $\langle  n^A(x) n^B(0) \rangle$ are of the form:
\begin{equation}
\langle  n^A(x) n^B(0) \rangle = \delta^{AB} D(r) \quad (A,B=1,2, \cdots, 8)
 .
\end{equation}
We have also checked that  one-point functions vanish: 
\begin{equation}
\langle  n^A(x)  \rangle = \pm 0.002 \simeq 0 \quad (A =1,2, \cdots, 8)
 .
\end{equation}
These results indicate  that the global $SU(3)$  \textbf{color symmetry} is preserved, that is to say, there is no specific direction in color space.  
This is expected, since the Yang-Mills theory should respect the global gauge symmetry, i.e., color symmetry, even after imposing the Landau gauge.

To obtain correlation functions of field variables, we need to fix the gauge and we have adopted the Landau gauge for the original Yang-Mills field $\mathscr{A}$ so that the global color symmetry is not broken.  This property is desirable to study color confinement, but it is lost in the MA gauge. 

\section{Gluon propagators and dominance}

\begin{figure}[t]
\includegraphics[scale=0.45]{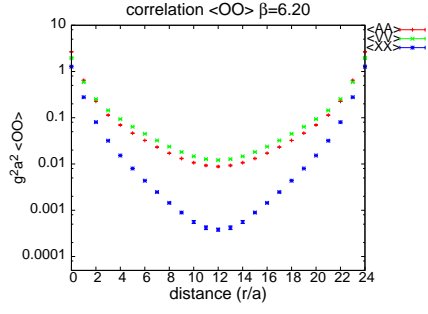}
\caption{\cite{KSSK11} 
Field correlators as functions of the distance $r:=|x|$ (from above to below)
$\langle  \mathscr{V}_\mu^A(x) \mathscr{V}_\mu^A(0) \rangle$, 
$\langle  \mathscr{A}_\mu^A(x) \mathscr{A}_\mu^A(0) \rangle$, 
and
$\langle  \mathscr{X}_\mu^A(x) \mathscr{X}_\mu^A(0) \rangle$.
}
\label{C35-fig:decomp-field-corr}
\end{figure}


We study the 2-point correlation functions (propagators) of the new variables and the original Yang-Mills field variables, which are defined by
\begin{equation}
D_{OO}(x-y):=\left\langle O_{\mu}^{A}(x)O_{\mu}^{A}(y)\right\rangle \text{ for} \  O_{\mu}^{A}(x ) \in \{\mathbb{V}^{A}_{x^{\prime},\mu}, \mathbb{X}^{A}
_{x^{\prime},\mu},\mathbb{A}^{A}_{x^{\prime},\mu}\},
\end{equation}
where an operator $\mathbb{O}_{\mu}(x) =O_{\mu}^{A}(x) T_A$  is defined by the linear type, e.g.,
$\mathbb{A}_{x^{\prime},\mu}:=(U_{x,\mu}-U_{x,\mu}^{\dag})/(2ig\epsilon)$ where $x^{\prime}$ means the mid-point of $x$ and $x+\epsilon \hat\mu$.
In order to calculate the propagators, we must impose a gauge fixing condition, and we have adopted the lattice Landau gauge (LLG).

Fig. \ref{C35-fig:decomp-field-corr} shows the 2-point correlation functions of new fields $\mathscr{V}$, $\mathscr{X}$, and original fields $\mathscr{A}$.
This result indicates the \textbf{infrared  dominance  of restricted correlation functions} $\langle  \mathscr{V}_\mu^A(x) \mathscr{V}_\mu^A(0) \rangle$ in the sense that the  correlator of the variable $\mathscr{V}$ behaves just like the correlator  $\langle  \mathscr{A}_\mu^A(x) \mathscr{A}_\mu^A(0) \rangle$ of the original variable $\mathscr{A}$ and dominates  in the long distance, while the correlator $\langle  \mathscr{X}_\mu^A(x) \mathscr{X}_\mu^A(0) \rangle$ of $SU(3)/U(2)$ variable $\mathscr{X}$  decreases quickly in the distance $r$.

\begin{figure}[ptb]
\includegraphics[
scale=0.6
]
{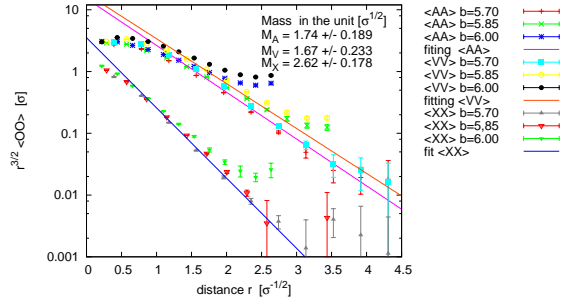} 
\caption{ \cite{SKKS13}
The rescaled correlation correlation functions $r^{3/2}\left\langle
O(r)O(0)\right\rangle$ for $O=\mathbb{A,V,X}$ for $24^{4}$ lattice with $\beta=5.7$, $5.85$, $6.0.$ 
The physical scale is set in units of the string tension $\sigma_{\text{phys}}^{1/2}$. 
The correlation functions have the profile of cosh type because of the periodic boundary condition, and hence we use  data within distance of the half size of lattice.
} 
\label{C35-fig.propagator}%
\end{figure}

For  $\mathscr{X}$,  at least, we can introduce a \textbf{gauge-invariant mass term}: 
\begin{equation}
\frac12 M_X^2 \mathscr{X}_\mu^A \mathscr{X}_\mu^A 
 ,
\end{equation} 
since $\mathscr{X}$ transforms like an adjoint matter field under the gauge transformation. 
In view of this fact, we fit the data of the contracted correlator $\langle  \mathscr{X}_\mu^A(x) \mathscr{X}_\mu^A(0) \rangle$ using the ``massive" propagator for large $r:=|x|$:
\begin{align}
 \langle  \mathscr{X}_\mu^A(x) \mathscr{X}_\mu^A(0) \rangle 
&=   \int \frac{d^4k}{(2\pi)^4} e^{ikx} \frac{3}{k^2+M_X^2} 
 \simeq  {\rm const.} \frac{e^{-M_X r}}{r^{3/2}} 
  .
\end{align}
In the similar way, we estimate the ``mass'' $M_{O}$ (i.e., the rate of exponential fall-off) from the propagator $D_{OO}(r)$ by using the Fourier transformation of the massive
propagator in the Euclidean space, which  behaves for large $M_{O}r$ as
\begin{equation}
D_{OO}(r)
=\left\langle O_{\mu}^{A}(x)O_{\mu}^{A}(y)\right\rangle
=\int\frac{d^{4}k}{(2\pi)^{4}}e^{ik(x-y)}\frac{3}{k^{2}+M_{O}^{2}%
}\simeq\frac{3\sqrt{M_{O}}}{2(2\pi)^{3/2}}\frac{e^{-M_{O}r}}{r^{3/2}}\text{
\ \ (}M_{O}r\gg1\text{)},
\end{equation}
and hence the scaled propagator $r^{3/2}D_{OO}(r)$ should be proportional to $\exp(-M_{O}r)$.

Fig.~\ref{C35-fig.propagator} shows the logarithmic plot of the scaled propagators $r^{3/2}D_{OO}(r)$ as a function of $r=|x-y|$, where the distance $r$ is drawn in units of the string tension $\sigma_{\text{phys}}$, and data of lattice spacing is taken from the Table I in Ref.\cite{Edward98}. 
The propagator $D_{VV}$ fall off slowly and has almost the same fall-off behavior as $D_{AA}$, while the $D_{XX}$ falls off quickly. Thus, from the viewpoint of the propagator, the $V$-field plays the dominant role in the deep infrared region or the long distance, while $X$-field is negligible in the long distance.
The rapid disappearance of $X$ contribution in the long distance is helpful to understand the difference of the profile of the flux tube in Fig.\ref{C35-fig:fluxtube}. 
In order to perform the parameter fitting of $M_{O}$ for $O=\{\mathbf{V}_{x^{\prime},\mu},\mathbf{A}_{x^{\prime},\mu}\}$, 
we use data in the region
$[2.0,4.5]$ and exclude the data near the midpoint of the lattice to eliminate the finite volume effect, while for $O=\mathbf{X}_{x^{\prime},\mu}$ we use the region $[1.0,3.5].$ 

Then the naively estimated ``mass" $M_X$ of $\mathscr{X}$ is 
\begin{equation}
 M_X = 2.409 \sqrt{\sigma_{\rm phys}} = 1.1 {\rm GeV} 
  .
\end{equation} 
We use $\sigma_{\text{phys}}=(440MeV)^2$ to obtain preliminary result:
\begin{equation}
M_{A}\simeq 0.76  \text{ GeV, \ \ \ }M_{V}\simeq 0.73\text{ GeV , \ \ \ }M_{X}%
\simeq 1.15 \text{ GeV,}
\end{equation}
which should be compared with result of the maximal option \cite{Shibata-lattice2007} in LLG, 
and also result of the Abelian projection in the MA gauge \cite{SAIIMT02,GIS12}.


\section{Conclusion and discussion}

We have combined a non-Abelian Stokes theorem for the Wilson loop operator \cite{Kondo08} and the new reformulations of the $SU(N)$ Yang-Mills theory on the lattice \cite{KSSMKI08,SKS10} according to a new viewpoint extended to the $SU(N)$ Yang-Mills theory  \cite{KSM08}, which provide  one with an efficient framework to study quark confinement from a viewpoint of the dual superconductor in the gauge-independent manner. 
 
We have presented the results of numerical simulations  of the lattice $SU(3)$ Yang-Mills theory \cite{KSSK11,SKKS13}, which support  the \textbf{non-Abelian dual superconductivity} for $SU(3)$ Yang-Mills theory proposed in \cite{KSSK11}.
We have shown that the restricted field extracted from the original $SU(3)$ Yang-Mills field plays a dominant role in confinement of quarks in the fundamental representation, i.e., the \textbf{restricted field dominance} in the (fundamental) string tension.
The restricted-field dominance was also confirmed for gluon propagators.

We have given numerical evidences that the non-Abelian magnetic monopoles defined in a gauge-invariant way  are dominant for confinement of fundamental quarks in $SU(3)$ Yang-Mills theory, i.e.,  \textbf{non-Abelian magnetic monopole dominance} in the (fundamental) string tension. 
By using the gauge invariant magnetic current $k$, we have extracted just the $U(1)$ part of  the maximal stability group $U(N-1) \simeq SU(N-1) \times U(1)$ for the non-Abelian magnetic monopole associated with quarks in the fundamental representation, which is consistent with the consideration of the Homotopy group. 
This $U(1)$ part is enough to extract the dominant part of the Wilson loop average.

In order to confirm the existence of the dual Meissner effect in $SU(3)$ Yang-Mills theory,   we have measured the gauge-invariant chromo field strength in the presence of a quark and an antiquark for both the original Yang-Mills field and the restricted  field. 
We have observed the dual Meissner effect in $SU(3)$ Yang-Mills theory: only the chromoelectric field exists in the flux tube connecting a quark and an antiquark and the associated magnetic-monopole current is induced around it. 
Moreover, we have determined the type of the non-Abelian dual superconductivity, i.e., \textbf{type I for the dual superconductivity of  $SU(3)$ Yang-Mills theory}, which should be compared with the border of type I and II for the dual superconductivity of the $SU(2)$ Yang-Mills theory.
These features are reproduced only from the restricted  part. 
 
In order to draw the definite conclusion on physical quantities in the continuum limit, e.g., the Ginzburg-Landau parameter, however, we must study the scaling of the data obtained in the numerical simulations. For this purpose, we need to accumulate more data at various choices for the gauge coupling on the lattices with different sizes. 
These features will be discussed in the future works. 
In the future, moreover, we hope to study the electric-current contribution to the Wilson loop average and the Abelian dominance and monopole dominance in the \textbf{adjoint Wilson loop} with the  possibilities of their connections to the \textbf{Casimir scaling} in the intermediate region and \textbf{string breaking} as a special case of \textbf{$N$-ality} in the asymptotic region.

For more preliminary results of numerical simulations, see \cite{Shibata-lattice2009} for magnetic monopoles of $SU(2)$,  \cite{Shibata-lattice2007} for the maximal option of $SU(3)$ and \cite{Shibata-lattice2010,Shibata-lattice2008} for the minimal one of $SU(3)$.
For more applications of the reformulation, see the recent review \cite{KKSS14}.

\section*{Acknowledgements}
This work is financially supported in part by Grant-in-Aid for Scientific Research (C) 24540252 from Japan Society for the Promotion of Science (JSPS).
This work is in part supported by the Large Scale Simulation Program 
No.09-15 (FY2009), No.T11-15 (FY2011), No.12/13-20 (FY2012-2013) and No.13/14-23 (FY2013-2014) of High Energy Accelerator Research Organization (KEK).



\bibliographystyle{aipproc}   


\begin{thebibliography}{99}






 





  


 
 


\bibitem{KSM08}
K.-I. Kondo, T. Shinohara and T. Murakami,
Prog.Theor. Phys. {\bf 120}, 1
(2008).
arXiv:0803.0176 [hep-th]


\bibitem{Cho80c}
Y.M. Cho,
Phys. Rev. Lett. {\bf 44}, 1115
(1980).


\bibitem{FN99a}
L. Faddeev and A.J. Niemi,
Phys. Lett. B {\bf 449}, 214
(1999).
[hep-th/9812090]
%
Phys. Lett. B {\bf 464}, 90
(1999).
[hep-th/9907180]


\bibitem{Cho80}
  Y.M. Cho,
Phys. Rev. D{\bf 21}, 1080
 (1980);
Phys. Rev. D{\bf 23}, 2415
(1981). 


\bibitem{DG79}
  Y.S. Duan and M.L. Ge, 
Sinica Sci., {\bf 11}, 1072
(1979). 


\bibitem{FN99} 
L. Faddeev and A.J. Niemi,
Phys. Rev. Lett. {\bf 82}, 1624
(1999).
 [hep-th/9807069],


\bibitem{Shabanov99}
  S.V. Shabanov,
Phys. Lett. B {\bf 458}, 322
 (1999).
[hep-th/9903223]
Phys. Lett. B {\bf 463}, 263
 (1999).
[hep-th/9907182]


\bibitem{KMS06}
  K.-I. Kondo, T. Murakami and T. Shinohara,
Prog. Theor. Phys. {\bf 115}, 201
(2006). 
[hep-th/0504107] 


\bibitem{KMS05}
  K.-I. Kondo, T. Murakami and T. Shinohara,
Eur. Phys. J. C{\bf 42}, 475
(2005).
[hep-th/0504198]


\bibitem{Kondo06}
K.-I. Kondo,
Phys. Rev. D{\bf 74}, 125003 (2006). 
[hep-th/0609166] 


\bibitem{KKMSSI06}
  S. Kato, K.-I. Kondo, T. Murakami, A. Shibata, T. Shinohara and S. Ito,
Phys. Lett. B{\bf 632}, 326
 (2006).
[hep-lat/0509069] 




\bibitem{IKKMSS07}
  S. Ito, S. Kato, K.-I. Kondo, T. Murakami, A. Shibata and T. Shinohara,  
Phys. Lett. B{\bf 645}, 67
(2007).  
[hep-lat/0604016] 


\bibitem{SKKMSI07}
A. Shibata, S. Kato, K.-I. Kondo, T. Murakami, T. Shinohara and  S. Ito,
Phys.Lett. B{\bf 653}, 101
(2007). 
arXiv:0706.2529 [hep-lat]

 
\bibitem{KKS14}
S. Kato, K.-I. Kondo, and  A. Shibata, 
arXiv:1407.2808 [hep-lat],


\bibitem{KKSS14}
K.-I. Kondo, S. Kato,  A. Shibata, and T. Shinohara, 
e-Print: arXiv:1409.1599 [hep-th].

 
\bibitem{DP89}
  D. Diakonov and V. Petrov, 
  Phys. Lett. B{\bf 224}, 131
 (1989).


\bibitem{Kondo98b}
  K.-I. Kondo,
  Phys. Rev. D{\bf 58}, 105016 (1998).
[hep-th/9805153]
  
  
\bibitem{KT99}
  K.-I. Kondo and Y. Taira,
Mod. Phys. Lett. A{\bf 15}, 367
(2000). 
[hep-th/9906129]
\\
K.-I. Kondo and Y. Taira,
Prog. Theor. Phys. {\bf 104}, 1189
 (2000).
[hep-th/9911242]


\bibitem{Kondo99Lattice99}
  K.-I. Kondo and Y. Taira,
Nucl. Phys. Proc. Suppl. {\bf 83}, 497
 (2000).


\bibitem{Kondo08}
  K.-I. Kondo,
Phys. Rev. D{\bf 77}, 085029 (2008).
arXiv:0801.1274 [hep-th]



\bibitem{KSSMKI08}
K.-I. Kondo, A. Shibata, T. Shinohara, T. Murakami, S. Kato and  S. Ito,
Phys. Lett. B{\bf 669}, 107
(2008). 
arXiv:0803.2451[hep-lat]


\bibitem{SKS10}
A. Shibata, K.-I. Kondo and T. Shinohara,
Phys. Lett. B{\bf 691}, 91
(2010).
arXiv:0911.5294 [hep-lat].


\bibitem{KS08}
 K.-I. Kondo and A. Shibata,
arXiv:0801.4203 [hep-th].


\bibitem{KSSK11}
 K.-I. Kondo, A. Shibata, T. Shinohara, and S. Kato,
Phys. Rev. D{\bf 83}, 114016 (2011). 
arXiv:1007.2696 [hep-th] 


\bibitem{SKKS13}
A. Shibata, K.-I. Kondo, S. Kato and T. Shinohara, 
Phys. Rev. D{\bf 87}, 054011 (2013).
arXiv:1212.6512 [hep-lat]







\bibitem{Clem75}
J.R. Clem,
J. Low. Temp. Phys. {\bf 18}, 427 (1975).


\bibitem{Cea:2012qw} 
P.~Cea, L.~Cosmai and A.~Papa,
Phys. Rev. D {\bf 86}, 054501 (2012).
arXiv:1208.1362 [hep-lat].


\bibitem {Edward98}
R.G. Edwards, U.M. Heller and T.R. Klassen, 
Phys. Rev. Lett. {\bf 80}, 3448--3451 (1998).


\bibitem{Matsubara:1993nq} 
Y.~Matsubara, S.~Ejiri and T.~Suzuki,
Nucl.\ Phys.\ Proc.\ Suppl.\  {\bf 34}, 176 (1994).
[hep-lat/9311061].


\bibitem{Chernodub:2005gz} 
M.~N.~Chernodub, K.~Ishiguro, Y.~Mori, Y.~Nakamura, M.~I.~Polikarpov, T.~Sekido, T.~Suzuki and V.~I.~Zakharov,
Phys.\ Rev.\ D {\bf 72}, 074505 (2005).
[hep-lat/0508004].


\bibitem{Suzuki:2009xy} 
T.~Suzuki, M.~Hasegawa, K.~Ishiguro, Y.~Koma and T.~Sekido,
Phys.\ Rev.\ D {\bf 80}, 054504 (2009).
[arXiv:0907.0583 [hep-lat]].


\bibitem{Cardoso:2010kw} 
N.~Cardoso, M.~Cardoso and P.~Bicudo,
arXiv:1004.0166 [hep-lat].


\bibitem{Shibata-lattice2007}
A. Shibata, S. Kato, K.-I. Kondo, T. Murakami, T. Shinohara, and S. Ito, 
PoS(LATTICE-2007)331, 
arXiv:0710.3221 [hep-lat] 


\bibitem{SAIIMT02} 
H. Suganuma, K. Amemiya, H. Ichie, N. Ishii, H. Matsufuru and T.T. Takahashi,
Nucl. Phys. B (Proc. Suppl.) {\bf 106}, 679--681 (2002).
[hep-lat/0407016], 


\bibitem{GIS12}
S. Gongyo, T. Iritani, and H. Suganuma, 
Phys. Rev. D{\bf 86},  094018 (2012).
e-Print: arXiv:1207.4377 [hep-lat] 
\\
S. Gongyo and H. Suganuma,
Phys. Rev. D{\bf 87},  074506 (2013). 
e-Print: arXiv:1302.6181 [hep-lat]


\bibitem{Shibata-lattice2008}
A. Shibata, K.-I. Kondo, S. Kato, S. Ito, T. Shinohara, and T. Murakami,
PoS LATTICE2008:268,2008. 
arXiv:0810.0956 [hep-lat], 


\bibitem{Shibata-lattice2009}
A. Shibata, K.-I. Kondo,  S. Kato, S. Ito, T. Shinohara, and N. Fukui,
PoS LATTICE2009:232,2009. 
arXiv:0911.4533 [hep-lat] 


\bibitem{Shibata-lattice2010}
A. Shibata, K.-I. Kondo, S. Kato, and T. Shinohara,
PoS LATTICE2010:286,2010. 


\bibitem{GMO90} 
A.~Di Giacomo, M.~Maggiore and S.~Olejnik,
Nucl.\ Phys.\ B{\bf 347}, 441 (1990). 
\\
%
A.~Di Giacomo, M.~Maggiore and S.~Olejnik,
Phys.\ Lett.\ B{\bf 236}, 199 (1990).


\end{thebibliography}

\IfFileExists{\jobname.bbl}{}
 {\typeout{}
  \typeout{******************************************}
  \typeout{** Please run "bibtex \jobname" to optain}
  \typeout{** the bibliography and then re-run LaTeX}
  \typeout{** twice to fix the references!}
  \typeout{******************************************}
  \typeout{}
 }



\end{document}